\documentclass[twocolumn,showpacs,preprintnumbers,amsmath,amssymb]{revtex4}
\usepackage{color}
\usepackage{bm, graphicx, amsmath}
\begin{document}

\title{Bell inequalities from group actions of single-generator groups}
\author{V. U\u{g}ur G\"{u}ney and Mark Hillery}
\affiliation{Department of Physics, Hunter College of the City University of New York, 695 Park Avenue, New York, NY 10065 USA}

\begin{abstract}
We study a method of generating Bell inequalities by using group actions of single-generator abelian groups.  Two parties, Alice and Bob, each make one of $M$ possible measurements on a system, with each measurement having $K$ possible outcomes.  The probabilities for the outcomes of these measurements are $P(a_{j}=k,b_{j^{\prime}}=k^{\prime})$, where $j,j^{\prime}\in \{ 1,2,\ldots M\}$ and $k,k^{\prime}\in \{ 0,1,\ldots K-1\}$.  The sums of some subsets of these probabilities have upper bounds when the probabilities result from a local, realistic theory that can be violated if the probabilities come from quantum mechanics.  In our case the subsets of probabilities are generated by a group action, in particular, a representation of a single-generator group acting on product states in a tensor-product Hilbert space.  We show how this works for several cases, including $M=2$, $K=3$, and general $M$, $K=2$.  We also discuss the resulting inequalities in terms of nonlocal games.
\end{abstract}


\maketitle

\section{Introduction}
Bell inequalities are conditions that must be satisfied by local, realistic theories \cite{bell}.  They did not attract much attention initially (see, however, \cite{clauser}-\cite{braunstein}), but the advent of quantum information, in particular the potential use of Bell inequalities in quantum cryptography \cite{ekert}, caused a surge in interest in them.  There are now many different Bell inequalities, and progress has been made in tabulating and classifying them (see, for example, \cite{vertesi}).  We cannot summarize what has become a large field or research, but two recent reviews \cite{brunner,liang} and a discussion of open questions \cite{gisin} give an excellent idea of the present state of the subject.

Bell inequalities come in many varieties, and they can be characterized by the number of parties making measurements, $N$, the number of measurement settings, $M$, and the number of possible outcomes for each measurement, $K$.  The original versions, the CHSH \cite{clauser} and CH \cite{ch} inequalities were for the case $N=M=K=2$.   Kaszlikowsi, et al.\ showed that by increasing the number of outcomes, $K$, one could more strongly violate local realism \cite{kaszlikowski}.  Full correlation Bell inequalities for the case  $M=K=2$ were fully characterized by Werner and Wolf, and, in addition, they were able to show how to derive maximum quantum violations of these inequalities \cite{werner2}.  Bell inequalities for the case $N=2$, $M=2$, and general $K$ were developed by Collins, et al.\ \cite{collins}, and this was generalized to the case of general $N$, $M=2$, and general $K$ by Son, et.\ al. \cite{son}.  

There is recent work on Bell inequalities,and non-contextuality inequalities by Cabello, Severini, and Winter that is related to what we will do here \cite{cabello}.  It is a very interesting approach that relates individual inequalities to graphs.  The terms in the inequalities are probabilities of events, and an event is a collection of measurement results, one measurement and one outcome of that measurement per party.  The vertices of the corresponding graph are events, and two vertices are connected by an edge if the events corresponding to them cannot be true simultaneously.  The independence number of the graph is an upper bound on the sum of the probabilities of the events when the probabilities are calculated from a local, realistic theory.  The Lovasz number of the graph is an upper bound for the sum of the probabilities of the events when they are calculated from quantum mechanics.

Like Cabello, \emph{et al.}, we want to look at Bell inequalities that are based on the sum of a subset of the basic probabilities that can be obtained by measuring the system, that is to say the probabilities of events.  For example, suppose there are two parties, Alice and Bob, and each can make $M$ different measurements, with each measurement having $K$ possible outcomes.  Let Alice's observables be $\{a_{j}|j=1,\ldots M\}$ and Bob's be $\{ b_{j}|j=1,\ldots M\}$.  The basic probabilities describing measurements by Alice and Bob can be written as $P(a_{j}=k, b_{j^{\prime}}=k^{\prime})$, which is just the probability that if Alice measures $a_{j}$ she gets $k$, and if Bob measures $b_{j^{\prime}}$, he gets $k^{\prime}$.  If these probabilities come from a local, realistic theory, there is an underlying joint probability distribution for all of the observables.  We shall often refer to this case as the classical case.  Classically the sum of subsets of these probabilities has an upper bound, which is a result of the fact that the probabilities all come from an underlying joint probability distribution.  This bound does not necessarily hold if the probabilities are derived from quantum mechanics, because the operators describing the observables may not commute, which rules out a joint distribution.  As we shall see, the result is that for some subsets of probabilities, when they are calculated quantum mechanically, their sum can violate the classical bound.

The sets of probabilities we consider will be generated by a group action \cite{rotman}.  If $G$ is a group and $X$ is a set, a group action is a function $\alpha: G\times X\rightarrow X$ such that $\alpha (e,x)=x$ and $\alpha (g,\alpha (h,x))=\alpha (gh,x)$.  Here $e,g,h\in G$ and $e$ is the identity element of the group.  The subset of $X$ given by $\{ \alpha (g,x)| g\in G\}$ is called the orbit of $x$.  Any two orbits are either distinct or identical, so the set of orbits forms a partition of $X$.  We will consider the simplest possible groups, abelian groups with a single generator.  If the generator is $g$, the group is just $G=\{ g^{r}|r=0,1,\ldots R-1\}$ where the order of the group is $R$ and $g^{R}=e$ (we also define $g^{0}=e$).  The set $X$ will be a Hilbert space and the group action will be given by a mapping of each element onto a unitary operator on that space, $U(g^{r}) =U^{r}(g)$.  In particular, the Hilbert space will be a product space, one factor for each party, and the operator $U(g)$ will yield a product state when it acts on a product state.  Therefore, the orbits will consist of product states, and these states will correspond to the eigenstates of local observables, and each of these eigenstates corresponds to a set of measurement results, one result for each party.

In some cases this procedure will lead to interesting correlated states, in particular, the states that produce the largest violations of the classical inequalities.  In order to see what can happen, let us look at the case $N,M=2$, so that Alice and Bob can each make two possible measurements.  Now suppose that Alice and Bob each choose at random an observable to measure, and they then measure it.  They then announce which observable they measured, and each, based on their measurement result, wishes to predict the measurement result of the other party.  For this to have a chance of working, their measurement results have to be correlated, and these correlations should be independent of the observable choice for each of them, because Alice did not know beforehand which observable Bob would choose, and vice versa.  The method we will describe here will produce quantum states with these kinds of correlations.

Finally, we would like to briefly compare the method described here to the graph-theory approach in \cite{cabello}.  In that approach one starts with a set of probabilities, and the graph they generate gives the classical and quantum bounds of the sum of that set of probabilities.  One then has to find measurements and a quantum system that correspond to those probabilities.  In our approach, we start with the quantum system and the measurements, which are generated by the group, and then one has to find the classical bound and see whether it can be violated.  The starting points of the methods are, therefore, different, and the methods can be viewed as complementary.

\section{Qubits}
In order to see how things work, let us start with the simplest case, $N=M=K=2$, which will result in the CH inequality.  The procedure we will outline here is easily generalized to more measurements and more measurement outcomes.  We begin by defining the translation operator, $T$, which has the action on the computational basis, which we shall also call the $z$ basis, 
\begin{equation}
\label{trans-qubit}
T|0\rangle = |1\rangle \hspace{5mm} T|1\rangle = |0\rangle .
\end{equation}
Note that in this case $T=\sigma_{x}$ so that it can be expressed as
\begin{equation}
T= |+x\rangle\langle +x| - |-x\rangle\langle -x| ,
\end{equation}
where $|\pm x\rangle = (|0\rangle \pm |1\rangle )/\sqrt{2}$.  We are interested in the operator
\begin{equation}
U= |+x\rangle\langle +x| +i |-x\rangle\langle -x| ,
\end{equation}
which has the property that $U^{2}=T$.  The condition $U^{2}=T$ does not determine $U$ uniquely, and we have made a particular choice, which, as we shall see, will lead to a violation of a classical inequality.  We see that
\begin{eqnarray}
U|0\rangle & = & \frac{1}{\sqrt{2}} (e^{i\pi /4}|0\rangle + e^{-i\pi /4}|1\rangle ) = |v_{0}\rangle \nonumber \\
U|1\rangle & = & \frac{1}{\sqrt{2}} (e^{-i\pi /4}|0\rangle + e^{i\pi /4}|1\rangle ) = |v_{1}\rangle .
\end{eqnarray}
We also find that $U|v_{0}\rangle = |1\rangle$ and $U|v_{1}\rangle = |0\rangle$.  We will consider the situation in which the two measurements that Alice and Bob can make are to measure in the computational basis, $\{ |0\rangle ,|1\rangle \}$ or in the $v$ basis, $\{ |v_{0}\rangle , |v_{1}\rangle \}$.

Now consider two qubits, and let the swap operator, $S$, act on the basis elements as $S|j\rangle |k\rangle = |k\rangle |j\rangle$, where $j,k=0,1$.  If we successively apply the operator $B=(U\otimes I)S$ to the state $|0\rangle |0\rangle$, we obtain the sequence of states
\begin{equation}
\label{qubit-sequence}
\begin{array}{ccc}
|0\rangle |0\rangle \rightarrow & |v_{0}\rangle |0\rangle \rightarrow & |v_{0}\rangle |v_{0}\rangle \rightarrow \\
|1\rangle |v_{0}\rangle \rightarrow & |1\rangle |1\rangle \rightarrow & |v_{1}\rangle |1\rangle  \rightarrow \\
|v_{1}\rangle |v_{1}\rangle \rightarrow & |0\rangle |v_{1}\rangle \rightarrow & |0\rangle |0\rangle 
\end{array}
\end{equation}
We see that what we have is a sequence of eight different states in which each state is  a member of either the computational or the $v$ basis.  If we take the inner products of these states with a specified two-qubit state $|\phi\rangle$ and square their magnitudes, we obtain the probabilities for the outcomes of measurements of this state by Alice and Bob in the computational and $v$ bases.  

Now suppose that we want to choose $|\phi\rangle$ to maximize the sum of this set of probabilities.  That means we want to maximize the quantity
\begin{equation}
\label{qubit-prob-sum}
\sum_{j=0}^{7} |\langle\phi |B^{j}|00\rangle |^{2} .
\end{equation}
This can be done by finding the largest eigenvalue of the operator
\begin{equation}
A= \sum_{j=0}^{7} B^{j}|00\rangle \langle 00| (B^{\dagger})^{j} .
\end{equation}
What we are going to show is that the eigenvectors of $B$ are eigenvectors of $A$, but before we do, notice what this implies.  If $|\phi\rangle$ is an eigenvector of $B$ all of the terms in the sum in Eq.\ (\ref{qubit-prob-sum}) are the same, which implies that the probabilities of the measurement outcomes in the set we are considering are all the same.

Now let us show that the eigenstates of $B$ are also eigenstates of $A$.  Let the eigenstates of $B$ be $\{ |u_{k}\rangle \}$, with corresponding eigenvalues $\lambda_{k}$.  We first note that $B^{2}= U\otimes U$, which implies that $B^{8}=I\otimes I$ since $U^{4}=I$.  This gives us that $\lambda_{k}^{8}=1$, so that $\lambda_{k}$ must be of the form $\exp (im\pi /4)$ for some integer $0\leq m \leq 7$.  We then have that
\begin{eqnarray}
\label{A-eigen}
A|u_{k}\rangle & = & \sum_{j=0}^{7} B^{j}|00\rangle \langle 00| (B^{\dagger})^{j}|u_{k}\rangle \nonumber \\
& = & \langle 00|u_{k}\rangle \sum_{j=0}^{7} B^{j} (\lambda_{k}^{\ast})^{j} \left( \sum_{l=0}^{3} |u_{l}\rangle\langle u_{l}|00\rangle \right) \nonumber \\
& = &  \langle 00|u_{k}\rangle \sum_{l=0}^{7} \left(  \sum_{j=0}^{3} (\lambda_{l} \lambda_{k}^{\ast})^{j} \right)  |u_{l}\rangle\langle u_{l}|00\rangle ) .
\end{eqnarray}
Because of the form of the eigenvalues, and assuming they are not degenerate (the case of degenerate eigenvalues will be considered later), the sum over $j$ is equal to $8\delta_{kl}$, giving us the final result that
\begin{equation}
A|u_{k}\rangle = 8 |\langle 00|u_{k}\rangle |^{2} |u_{k}\rangle .
\end{equation}
Therefore, what we need to find to maximize the sum in Eq.\ (\ref{qubit-prob-sum}) is the eigenstate of $B$ with the largest overlap with the state $|00\rangle$.

The next step is to find the eigenstates of $B$.  The eigenstates of $U$ are $|+x\rangle$ and $|-x\rangle$, with eigenvalues $1$ and $i$, respectively.  This immediately implies that
\begin{equation}
|u_{0}\rangle = |+x\rangle |+x\rangle \hspace{5mm} |u_{1}\rangle = |-x\rangle |-x\rangle ,
\end{equation}
are eigenstates of $B$, with eigenvalues of $1$ and $i$, respectively.  In order to find the remaining two eigenvectors we note that
\begin{equation}
B|+x\rangle |-x\rangle = i |-x\rangle |+x\rangle \hspace{5mm} B|-x\rangle |+x\rangle = |+x\rangle |-x\rangle ,
\end{equation}
From these equations we find the two remaining eigenstates
\begin{eqnarray}
|u_{2}\rangle & = & \frac{1}{\sqrt{2}}( |+x\rangle |-x\rangle + e^{i\pi /4} |-x\rangle |+x\rangle ) \nonumber \\
|u_{3}\rangle & = & \frac{1}{\sqrt{2}}( |+x\rangle |-x\rangle - e^{i\pi /4} |-x\rangle |+x\rangle ) ,
\end{eqnarray}
with eigenvalues of $e^{i\pi /4}$ and $e^{-i\pi /4}$, respectively.  The eigenstate with the largest overlap with $|00\rangle$ is $|u_{2}\rangle$, which is
\begin{equation}
|\langle 00|u_{2}\rangle |= \frac{1}{2}\left( 1+ \frac{1}{\sqrt{2}}\right)^{1/2} ,
\end{equation}
and this gives a maximum value for the sum in Eq.\ (\ref{qubit-prob-sum}) of $8 |\langle 00|u_{2}\rangle |^{2}=2 +\sqrt{2}$.

We now need to find a classical bound on the sum of these eight probabilities.  Let us let $a_{1}$ and $b_{1}$ correspond to measuring in the $z$ basis, and $a_{2}$ and $b_{2}$ correspond to measuring in the $v$ basis.  The state $|j\rangle$, where $j=0,1$, corresponds to $a_{1}$ or $b_{1}$ being equal to $j$, and the state $|v_{j}\rangle$ corresponds to $a_{2}$ or $b_{2}$ being equal to $j$.  We now assume that we have a joint distribution for all of the observables, $P(a_{1},a_{2};b_{1},b_{2})$, and we want to express the sum of the probabilities corresponding to the states in Eq.\ (\ref{qubit-sequence}) in terms of this distribution.  We note that
\begin{eqnarray}
P(a_{1}=0,b_{1}=0)+P(a_{1}=1,b_{1}=1)  \nonumber \\
=  \sum_{a_{2},b_{2}} \sum_{j=0}^{1} P(a_{1}=j,a_{2}; b_{1}=j,b_{2}) \nonumber \\
P(a_{2}=0,b_{1}=0)+P(a_{2}=1,b_{1}=1) \nonumber \\
 =  \sum_{a_{1},b_{2}} \sum_{j=0}^{1} P(a_{1},a_{2}=j; b_{1}=j,b_{2}) \nonumber \\
P(a_{1}=1,b_{2}=0)+P(a_{1}=0,b_{2}=1) \nonumber \\
 =  \sum_{a_{2},b_{1}} [ P(a_{1}=1,a_{2}; b_{1},b_{2}=0) \nonumber \\
 + P(a_{1}=0,a_{2};b_{1},b_{2}=1) ] \nonumber \\
P(a_{2}=0,b_{2}=0)+P(a_{2}=1,b_{2}=1) \nonumber \\ 
=  \sum_{a_{1},b_{1}} \sum_{j=0}^{1} P(a_{1},a_{2}=j; b_{1},b_{2}=j) . \nonumber \\ 
\end{eqnarray}
Adding all of these equations together, we find that the sum of the eight probabilities on the left-hand sides, the quantity for which we want to find an upper bound, can be expressed as a linear combination of the probabilities on the right-hand sides, $\sum_{a_{1},a_{2},b_{1},b_{2}} c_{a_{1},a_{2},b_{1},b_{2}} P(a_{1},a_{2},b_{1},b_{2})$, where the $c_{a_{1},a_{2},b_{1},b_{2}}$ are positive integers.  In order to maximize this expression, we just want to choose the joint distribution to be equal to one in the term with the largest value of $c_{a_{1},a_{2},b_{1},b_{2}}$, and this implies that the sum of the eight probabilities will be less than or equal to the maximum value of $c_{a_{1},a_{2},b_{1},b_{2}}$.  In this case we find that this is equal to $3$, and this is then the  classical bound.  As we saw, the quantum mechanical bound was $2 +\sqrt{2}$, which is larger, so the state $|u_{2}\rangle$ violates the classical bound.

Let us now examine the properties of the state $|u_{2}\rangle$.  Because Alice and Bob can each make two measurements and each measurement has two outcomes, there are $16$ probabilities that describe the results of Alice's and Bob's measurements.  Eight of these probabilities appear in the sum in Eq.\ (\ref{qubit-prob-sum}), and each of these probabilities is equal to $(2+\sqrt{2})/8\simeq 0.427$  We can express the correlations present in this state by means of a nonlocal game \cite{cleve}.  A referee sends a bit, $s$, to Alice and another, $t$, to Bob, with each bit equally likely to be $0$ or $1$.  Alice and Bob then each send a bit back to the referee.  Alice and Bob win the game if one of the following conditions is satisfied:
\begin{enumerate}
\item If $(s,t)$ is $(0,0)$, $(1,0)$, or $(1,1)$, then the bits that Alice and Bob send are the same.
\item If $(s,t)$ is $(0,1)$, then the bits that Alice and Bob send are different.
\end{enumerate}
These conditions can be summarized as $\bar{s}\wedge t= a + b$, where $\bar{s}$ is the negation of $s$, $a$ and $b$ are the bit values returned by Alice and Bob, respectively, and the addition on the right-hand side is modulo 2.  The maximum classical probability of winning this game is $3/4$, but we can do better with a quantum strategy.  Suppose Alice and Bob share the state $|u_{2}\rangle$.  Let a bit value of $0$ correspond to the $z$ basis and a bit value of $1$ correspond to the $v$ basis.  When the referee sends the bits $s$ and $t$ to Alice and Bob, they measure their qubits in the corresponding bases.  They then send the results of their measurements to the referee as their bits.  For each pair of bases $(z,z)$, $(v,z)$, and $(v,v)$, where we have listed Alice's basis first, the probability that they agree is $(2+\sqrt{2})/4$, and for the basis choice $(z,v)$, the probability that they disagree is $(2+\sqrt{2})/4$.  Therefore, the probability that they win the game is $(2+\sqrt{2})/4 \simeq 0.85$, which is larger than that of the classical strategy.   A second, and related way of viewing the correlations present in the state $|u_{2}\rangle$ is to suppose that Alice and Bob share the state, and each chooses at random whether to measure in the $z$ or $v$ basis.  They then announce their basis choices.  Alice wants to predict what the result of Bob's measurement is based on the result of her measurement.  If the basis choice is one of $(z,z)$, $(v,z)$, or $(v,v)$ her prediction should be the same as her result, and if it is $(z,v)$, her prediction should be the opposite of her result.  Her probability of success is $(2+\sqrt{2})/4$.

\section{Qudits and more than two measurements}
What we have done can be generalized to measurements with more than two outcomes and more than two measurements per party.  For now, we shall only consider two parties.  We first assume that each measurement has $d$ outcomes, so we are dealing with qudits, and the computational basis is now $\{|j\rangle\, | \, j=0,1,\ldots d-1\}$.  The translation operator now has the action $T|j\rangle = |j+1\rangle$, where the addition is modulo $d$, and we define an operator $U$ so that $U^{n}=T$.  We now define the $n$ measurement bases $\{ |v_{j}^{(m)}\rangle =\{  U^{m}|j\rangle \, | \, j=0, \ldots d-1\}$, where $m=0,1,\ldots n-1$.  We again define the operator $B=(U\otimes I)S$, where $S$ is the swap operator, on $\mathcal{C}^{d}\otimes \mathcal{C}^{d}$.  We have that $B^{2}=U\otimes U$ and $T^{d}=I$, and these imply that $B^{2nd}=I\otimes I$.

Repeated application of $B$ to the vector $|0\rangle |0\rangle$ generates the sequence of states
\begin{equation}
\begin{array}{ccc}
|0\rangle |0\rangle\rightarrow & |v_{0}^{(1)}\rangle |0\rangle \rightarrow & |v_{0}^{(1)}\rangle  |v_{0}^{(1)}\rangle \rightarrow   \\
 v_{0}^{(2)}\rangle |v_{0}^{(1)} \rangle \rightarrow & |v_{0}^{(2)}\rangle |v_{0}^{(2)}\rangle \rightarrow & \ldots \\
|1\rangle |1\rangle \rightarrow & \ldots & |v_{d-1}^{(n-1)}\rangle |v_{d-1}^{(n-1)}\rangle \rightarrow \\
|0\rangle  |v_{d-1}^{(n-1)}\rangle\rightarrow & |0\rangle |0\rangle &  
\end{array}
\end{equation}
The elements of this sequence are of the form $B^{j}|00\rangle$ for $0\leq j \leq 2nd$.  We need to specify exactly which states appear in this sequence.  The state $|v^{(m_{1})}_{j}\rangle |v^{(m_{2})}_{k}\rangle$ will appear if one of the following conditions is satisfied.
\begin{enumerate}
\item If $m_{1}=m_{2}$ and $j=k$, i.e.\ states of the form $|v_{j}^{(m)}\rangle |v_{j}^{(m)}\rangle$.
\item If $m_{1}=m_{2}+1$ and $j=k$ for $0\leq m_{2}\leq n-2$, i.e.\ states of the form $|v_{j}^{(m+1)}\rangle |v_{j}^{(m)}\rangle$ for $0\leq m \leq n-2$.
\item If $m_{1}=0$, $m_{2}=n-1$, and $j=k+1$, i.e.\ states of the form $|v_{j+1}^{(0)}\rangle |v_{j}^{(n-1)}\rangle$, where the addition in the subscript is modulo $d$.
\end{enumerate}
There are $2nd$ distinct vectors in the sequence, and in the case $n=2$, each of the four possible pairings of the two bases will appear.

We are again interested in finding a vector, $|\phi\rangle$, that maximizes the sum of the probabilities
\begin{equation}
\sum_{j=0}^{2nd-1} |\langle\phi |B^{j}|00\rangle |^{2} ,
\end{equation}
which can be done by finding the largest eigenvalue of the operator 
\begin{equation}
A= \sum_{j=0}^{2nd-1} B^{j}|00\rangle\langle 00|(B^{\dagger})^{j} .
\end{equation}
As before, we can show that the eigenstates of $B$ can be used to find the eigenstates of $A$.  In fact, the operators commute so they are simultaneously diagonalizable.  We first note that the equation $B^{2nd}=I\otimes I$ implies that the eigenvalues of $B$ are of the form $e^{2\pi im/(2nd)}$ for some $0\leq m \leq 2nd-1$.  Following the argument in Eq.\ (\ref{A-eigen}) we see that if $|u_{k}\rangle$ is an eigenstate of $B$ with eigenvalue $\lambda_{k}$, then
\begin{equation}
A|u_{k}\rangle =  \langle 00|u_{k}\rangle \sum_{l=0}^{d^{2}-1} \left(  \sum_{j=0}^{2nd-1} (\lambda_{l} \lambda_{k}^{\ast})^{j} \right)  |u_{l}\rangle\langle u_{l}|00\rangle ) .
\end{equation}
The sum over $j$ yields zero unless $\lambda_{l}=\lambda_{k}$, so that we have
\begin{equation}
\label{degen}
A|u_{k}\rangle = 2nd\langle 00|u_{k}\rangle \sum_{ \{ l\, |\lambda_{l}=\lambda_{k} \} } |u_{l}\rangle \langle u_{l}|00\rangle .
\end{equation}
If the eigenvalue $\lambda_{k}$ is non-degenerate, then $|u_{k}\rangle$ is an eigenstate of $A$ with an eigenvalue of $2nd |\langle00|u_{k}\rangle |^{2}$.

 If the eigenvalue $\lambda_{k}$ is degenerate, we have to diagonalize $A$ within the subspace spanned by the eigenvectors with eigenvalue $\lambda_{k}$.  Let us denote the eigenvectors of $B$ with eigenvalue $\lambda_{k}$ by $\{ |u_{kl}\rangle \}$ and the dimension of the space spanned by then by $d_{k}$.  From Eq.\ (\ref{degen}) we see that $A$ maps all of the vectors $|u_{kl}\rangle$ onto the same vector 
\begin{equation}
|X_{k}\rangle = \sum_{l^{\prime}}|u_{kl^{\prime}}\rangle \langle u_{kl^{\prime}}|00\rangle .
\end{equation}
This vector is an eigenvector of $A$, and we have that
\begin{eqnarray}
A|X_{k}\rangle & = & \sum_{l^{\prime}}\left[ 2nd\langle 00|v_{l^{\prime}}\rangle \sum_{l^{\prime\prime}}|u_{kl^{\prime\prime}}\rangle\langle u_{kl^{\prime\prime}}|00\rangle \right] \langle u_{kl^{\prime}}|00\rangle \nonumber \\
 & = & \left( 2nd\sum_{l^{\prime}}|\langle 00|u_{kl^{\prime}}\rangle |^{2}\right) |X_{k}\rangle ,
\end{eqnarray}
so that its eigenvalue is just $\left( 2nd\sum_{l^{\prime}}|\langle 00|u_{kl^{\prime}}\rangle |^{2}\right)$.  The other eigenvalues of $A$ in the span of $\{ |u_{kl}\rangle \}$ are zero.  This can be seen by noting that 
\begin{equation}
A|u_{kl}\rangle = 2nd\langle 00|u_{kl}\rangle |X_{k}\rangle ,
\end{equation}
which implies that the $d_{k}-1$ linearly independent vectors, $\{ \langle 00|u_{kl}\rangle |u_{k1}\rangle - \langle 00|u_{k1}\rangle |u_{kl}\rangle \, |\, l=2, \ldots d_{k} \}$ all have an eigenvalue of $0$.

Now let us find the eigenstates of $B$.  Let us denote the eigenstates of $U$ by $|w_{j}\rangle$, with corresponding eigenvalues $\lambda_{j}$.  First, we note that $|w_{j}\rangle \otimes |w_{j}\rangle$ is an eigenstate of $B$ with eigenvalue $\lambda_{j}$.  Other eigenvectors will lie in $2\times 2$ blocks spanned by the vectors $|w_{j}\rangle$ and $|w_{k}\rangle$.  This can be seen by noting that
\begin{eqnarray}
B(|w_{j}\rangle |w_{k}\rangle ) & = & \lambda_{k} |w_{k}\rangle |w_{j}\rangle \nonumber \\
B(|w_{k}\rangle |w_{j}\rangle ) & = & \lambda_{j}|w_{j}\rangle |w_{k}\rangle ,
\end{eqnarray}
These equations imply that if $\lambda$ is an eigenvalue of $B$ with an eigenvector lying in the space spanned by $|w_{j}\rangle |w_{k}\rangle $ and $|w_{k}\rangle |w_{j}\rangle$, then it satisfies $\lambda^{2}=\lambda_{j}\lambda_{k}$, so that $\lambda = \pm \sqrt{\lambda_{j}\lambda_{k}}$.  The corresponding eigenvectors are
\begin{equation}
\frac{1}{\sqrt{2}}\left( |w_{j}\rangle |w_{k}\rangle \pm \frac{\sqrt{\lambda_{j}\lambda_{k}}}{\lambda_{j}} |w_{k}\rangle |w_{j}\rangle \right) .
\end{equation} 

\section{Two examples}
\subsection{Two qutrits, two three-valued measurements}
Now consider two qutrits, with each qutrit Hilbert space being spanned by the orthonormal basis $\{ |0\rangle , |1\rangle , |2\rangle \}$.  The eigenstates of the translation operator are 
\begin{equation}
|w_{j}\rangle  =  \frac{1}{\sqrt{3}} \sum_{k=0}^{2} e^{2\pi ijk/3} |k\rangle ,
\end{equation}
with eigenvalues $e^{-2\pi ij /3}$, for $j=0,1,2$.  We then have that 
\begin{equation}
T=\sum_{j=0}^{2} e^{-2\pi ij/3}|w_{j}\rangle\langle w_{j}| ,
\end{equation}
so for $U$ we choose
\begin{equation}
U= |w_{0}\rangle\langle w_{0}| + e^{-i\pi /3}|w_{1}\rangle\langle w_{1}| + e^{i\pi /3}|w_{2}\rangle\langle w_{2}| .
\end{equation}
The basis $|v_{j}\rangle = U|j\rangle$, $j=0,1,2$, is given explicitly by
\begin{eqnarray}
|v_{0}\rangle & = & \frac{1}{3}(2|0\rangle + 2|1\rangle -|2\rangle ) \nonumber \\
|v_{1}\rangle & = & \frac{1}{3}(-|0\rangle + 2|1\rangle + 2|2\rangle ) \nonumber \\
|v_{2}\rangle & = & \frac{1}{2}(2|0\rangle - |1\rangle + 2|2\rangle ) .
\end{eqnarray}

The eigenvalues of $B$ are $\pm 1$, $e^{i\pi /3}$, $e^{-i\pi /3}$, $\pm e^{i\pi /6}$, and $\pm e^{-i\pi /6}$.  The eigenvalue $1$ is the only one that is degenerate, and the eigenstates corresponding to it are the ones that lead to the eigenstate of $A$ with the largest eigenvalue.  The eigenstates of $B$ with eigenvalue $1$ are $|w_{0}\rangle|w_{0}\rangle$ and 
\begin{equation}
|w_{12}\rangle =\frac{1}{\sqrt{2}} ( |w_{1}\rangle |w_{2}\rangle + e^{i\pi /3}|w_{2}\rangle |w_{1}\rangle ) .
\end{equation}
The nonzero eigenvalue of $A$ in the space spanned by these two vectors is 
\begin{equation}
12 ( |\langle 0|w_{0}\rangle |^{4} + |\langle 00|w_{12}\rangle |^{2}) = \frac{10}{3} ,
\end{equation}
and the corresponding normalized eigenvector is
\begin{eqnarray}
|X_{1}\rangle & = & \sqrt{\frac{2}{5}} \left[ |w_{0}\rangle |w_{0}\rangle + \frac{1}{\sqrt{2}} (1+e^{-i\pi /3}) |w_{12}\rangle \right] \nonumber \\
 & = & \sqrt{\frac{2}{5}}\left[ \frac{5}{6}(|00\rangle + |11\rangle + |22\rangle ) \right. \nonumber \\
 & & + \frac{1}{3}(|10\rangle + |02\rangle +|21\rangle )  \nonumber \\
& & \left.  -\frac{1}{6}(|01\rangle + |20\rangle + |12\rangle ) \right] .
\end{eqnarray}

Now, suppose Alice and Bob share the state $|X_{1}\rangle$, and suppose that Alice measures either  $a_{1}$ or $a_{2}$, and Bob measures either $b_{1}$ or $b_{2}$.  Each of these observables takes the values $0$, $1$, or $2$.  For $a_{1}$ and $b_{1}$, the eigenstates are $\{ |j\rangle \, | \, j=0,1,2\}$, with the eigenstate $|j\rangle$ corresponding to the eigenvalue $j$, and for $a_{2}$ and $b_{2}$, the eigenstates are $\{ |v_{j}\rangle \, | \, j=0,1,2\}$, with the eigenstate $|v_{j}\rangle$ corresponding to the eigenvalue $j$.  The probabilities $P(a_{1}=j,b_{1}=j)$, $P(a_{2}=j,b_{1}=j)$, $P(a_{2}=j,b_{2}=j)$, for $j=0,1,2$, and the probabilities $P(a_{1}=0,b_{2}=2)$, $P(a_{1}=1,b_{2}=0)$, $P(a_{1}=2,b_{2}=1)$, are, by construction, all the same and are equal to $5/18$ in the state $|X_{1}\rangle$.  Their sum is just the eigenvalue of $A$ corresponding to $|X_{1}\rangle$, which, as we saw, is $10/3$.  If these probabilities came from a joint distribution for all four observables, the maximum value their sum can have is $3$, so we obtain a quantum violation of the classical inequality.

This violation can, as before, be phrased in terms of a nonlocal game.  The game is almost the same as before, except that now Alice and Bob return one of three possible answers instead of two.  Each is sent a bit, $s$ to Alice and $t$ to Bob, and they win if $(s,t)$ is $(0,0)$, $(1,0)$, or $(1,1)$ and they return the same answer, or if $(s,t)=(0,1)$ and they return $(0,2)$, $(1,0)$ or $(2,1)$.  This winning condition can be summarized as $\bar{s}\wedge t= a - b$, where the subtraction is modulo 3.  The maximum classical probability of winning this game is again $3/4$.  Quantum mechanically, Alice and Bob share $|X_{1}\rangle$, Alice measures $a_{1+s}$ and Bob measures $b_{1+t}$, and they report their results.  With this strategy their probability of winning is $5/6$.

\subsection{Two qubits, three, or more, two-valued measurements}
We will now return to the case of two qubits, but consider the situation in which there are three measurement bases for each qubit rather than two.  The action of the translation operator is again given by Eq.\ (\ref{trans-qubit}), and we now want an operator, $U$, such that $U^{3}=T$.  We shall choose
\begin{equation}
U=|+x\rangle\langle +x| + e^{i\pi /3}|-x\rangle\langle -x| ,
\end{equation}
and define two new bases $\{ |v_{j}^{(k)}\rangle = U^{k}|j\rangle\, | \, j=0,1 \}$ for $k=1,2$ (we will sometimes denote the elements of the computational basis as $|v_{j}^{(0)}\rangle$, $j=0,1$).  Applying the operator $B$ to the state $|0\rangle |0\rangle$, we obtain the sequence of states
\begin{eqnarray}
\label{seq-2qb-3m}
|0\rangle |0\rangle \rightarrow |v_{0}^{(1)}\rangle |0\rangle \rightarrow |v_{0}^{(1)}\rangle |v_{0}^{(1)}\rangle \rightarrow |v_{0}^{(2)}\rangle |v_{0}^{(1)}\rangle \rightarrow \nonumber \\
|v_{0}^{(2)}\rangle |v_{0}^{(2)}\rangle \rightarrow |1\rangle |v_{0}^{(2)}\rangle \rightarrow |1\rangle |1\rangle \rightarrow |v_{1}^{(1)}\rangle |1\rangle \rightarrow \nonumber \\
|v_{1}^{(1)}\rangle |v_{1}^{(1)}\rangle \rightarrow |v_{1}^{(2)}\rangle |v_{1}^{(1)}\rangle \rightarrow  |v_{1}^{(2)}\rangle |v_{1}^{(2)}\rangle \rightarrow  |0\rangle |v_{1}^{(2)}\rangle .
\end{eqnarray}
The eigenvalues of $U$ are clearly $1$ and $e^{i\pi /3}$, which implies that the eigenvalues of $B$ are $1$, $e^{i\pi /3}$, and $\pm e^{i\pi /6}$, and none of these eigenvalues is degenerate.  The eigenstate corresponding to $e^{i\pi /6}$, which is given by
\begin{eqnarray}
|X\rangle & = & \frac{1}{\sqrt{2}}( |+x\rangle |-x\rangle + e^{i\pi /6} |-x\rangle |+x\rangle ) \nonumber \\
 & = & \frac{1}{2\sqrt{2}}[ (1+e^{i\pi /6})(|0\rangle |0\rangle - |1\rangle |1\rangle ) \nonumber \\
&& + (1-e^{i\pi /6})(|1\rangle |0\rangle - |0\rangle |1\rangle ) ] .
\end{eqnarray}
yields the largest eigenvalue for $A$, which is 
\begin{equation}
\label{2qb-3m-max}
12 |\langle X|00\rangle |^{2} = \frac{3}{2}(2+\sqrt{3}) .
\end{equation}

We now return to Alice and Bob, who now can perform one of three measurements on their respective qubits.  Alice can measure one of the three observables $a_{k}=|v_{1}^{(k)}\rangle\langle v_{1}^{(k)}|$ on her qubit and Bob can measure one of the three observables $b_{k}=|v_{1}^{(k)}\rangle\langle v_{1}^{(k)}|$ on his qubit, where $k=0,1,2$.  We now look at the probabilities corresponding to the sequence of states in Eq.\ (\ref{seq-2qb-3m}), e.g.\ the state $|0\rangle |0\rangle$ corresponds to $P(a_{0}=0,b_{0}=0)$ and $|1\rangle |v_{0}^{(2)}\rangle$ corresponds to $P(a_{0}=1,b_{2}=0)$.  If we sum these probabilities for the state $|X\rangle$, we obtain the result in Eq.\ (\ref{2qb-3m-max}), which is approximately $5.598$.  If all of these probabilities come from a joint distribution for all six observables, the largest value this sum can have is $5$.  Therefore, the quantum result violates the classical bound.

This result can also be phrased in terms of a nonlocal game.  Alice is sent $s$ and Bob is sent $t$, where $s$ and $t$ take values in the set $\{ 0,1,2\}$, $s-t=0,1$ modulo 3, with all allowed choices being equally likely.  They each have to return a bit.  They win if $(s,t)$ is $(0,0)$, $(1,0)$, $(1,1)$, $(2,1)$, or $(2,2)$ and they return the same bit value, or if $(s,t)=(0,2)$ and they return opposite bit values.  The maximum classical probability of winning is $5/6 \simeq .833$ (this can be achieved by always returning $(0,0)$).  If Alice and Bob share the state $|X\rangle$, let $(s,t)$ determine their choice of observable to measure, and return their measurement results, their probability of winning is $(2+\sqrt{3})/4 \simeq 0.933$, which is greater than the classical result.

This case can easily be generalized to $M$ measurements with each measurement having two possible outcomes.  In this case we have that
\begin{equation}
U=|+x\rangle\langle +x| + e^{i\pi /M}|-x\rangle\langle -x| ,
\end{equation}
and we have $M$ bases $\{ |v_{j}^{(k)}=U^{k}|j\rangle \, | \, l=0,1\}$ for $k=0,1,\ldots M-1$.  The eigenvalues of $U$ are $1$ and $e^{i\pi /M}$, which implies that the eigenvalues for $B$ are $1,\, e^{i\pi /M}$, and $\pm e^{i\pi /2M}$, none of which are degenerate.  The eigenvalue $e^{i\pi /2M}$, whose eigenvector is
\begin{equation}
\frac{1}{\sqrt{2}}(|+x\rangle |-x\rangle +e^{i\pi /2M}|-x\rangle|+x\rangle ),
\end{equation}
yields the largest eigenvalue of $A$, which is 
\begin{equation}
\label{M-meas}
M \left[ 1 + \cos \left(\frac{\pi}{2M}\right) \right] .
\end{equation}

Defining, as before, $a_{k}=|v_{1}^{(k)}\rangle\langle v_{1}^{(k)}|$ on Alice's qubit and $b_{k}=|v_{1}^{(k)}\rangle\langle v_{1}^{(k)}|$ on Bob's, for $k=0,1,\ldots M-1$, the sum of probabilities we are considering is
\begin{eqnarray}
R & = & \sum_{j=0,1} \left[ \sum_{k=0}^{M-1} P(a_{k}=j,b_{k}=j) \right. \nonumber \\
& & \left. + \sum_{k=0}^{M-2} P(a_{k+1}=j,b_{k}=j) \right] \nonumber \\
& & + P(a_{0}=0,b_{M-1}=1) \nonumber \\
& & + P(a_{0}=1,b_{M-1}=0) .
\end{eqnarray}
If these probabilities come from a joint distribution, $P(a_{0},a_{1},\ldots a_{M-1};b_{0},\ldots b_{M-1})$, then the probability with all variables set equal to zero will contribute to $2M-1$ of the terms in the above sum, and we cannot have a larger contribution than this.  This gives us a classical bound of $2M-1$ for $R$, while the quantum bound is given by Eq.\ (\ref{M-meas}).  The fact that the function $f(x)=(1/2)\cos (\pi x) + x \geq 1/2$ for $0<x\leq 1/2$ ($f(0)=f(1/2)=1/2$ and $f(x)$ is concave downward) implies that the quantum bound is greater than the classical one for $M\geq 2$.

Phrasing the result in terms of a nonlocal game, we have that Alice is sent $s$ and Bob $t$, where $s,t\in \{ 0,1,\ldots M-1\}$ and $s-t =0,1$ mod $M$.  Alice and Bob each send a bit.  They win if the bits are opposite when $(s,t)=(0,M-1)$ and if the bits agree for all other allowed choices of $(s,t)$.  The maximum classical probability of winning, which can be achieved by Alice and Bob always returning bit values of $0$, is $1-(1/(2M))$.  The quantum winning probability is achieved by Alice and Bob measuring their shared quantum state, with $s$ and $t$ dictating which variables to measure, and Alice and Bob returning their measurement results as the bit values.  This results in a winning probability of 
\begin{equation}
\frac{1}{2} \left[ 1+ \cos \left( \frac{\pi}{2M}\right) \right] \simeq 1-\frac{\pi^{2}}{16M^{2}} ,
\end{equation}
where the last expression is valid in the limit of large $M$.  In this limit, we see that the quantum winning probability approaches one faster than does the classical probability.  A related example of this type appears in \cite{liang}. 

We have checked the inequalities in this section agains those in the Faacets database of Bell inequalities and did not find any matches \cite{faacets}.

\section{Numerical results}
Now that we have a method to generate new Bell inequalities, we can apply it to more complicated situations, which we shall now do numerically.  Let us consider the case in which there are two parties each having two measurement choices, but the number of outcomes of each measurement is $d$ (we have already examined the cases $d=2,3$ in Sec.\ II and IV).  We find that \newline
\begin{center}
\begin{tabular}{c|c|c|c|c} $d$ & $Q_{s}$ & $C_{s}$ & $p$ & $I_{ab}$ \\ \hline 2 & 3.4142 & 3 & 0.8536 & 0.3991 \\ 3 & 3.3333 & 3 & 0.8333 & 0.8146 \\ 4 & 3.3066 & 3 & 0.8266 & 1.1482 \\ 5 & 3.2944 & 3 & 0.8236 & 1.4223
\end{tabular} 
\newline
\end{center}
where $Q_{s}$ is the sum of the probabilities calculated from quantum mechanics, $C_{s}$ is the upper bound on the sum of the corresponding classical probabilities, $p$ is the probability that Alice can correctly predict the result of Bob's measurement, and $I_{ab}$ is the mutual information between Alice and Bob.  These quantities require some explanation.  Application of the operator $B$ to the state $|0\rangle |0\rangle$ generates $4d$ states, which correspond to $4d$ probabilities (see, for example, Eq.\ \ref{qubit-sequence}), and we shall call this set of probabilities $S_{p}$.  The sum of these probabilities produced by the state that corresponds to the largest eigenvalue of the operator $A$ is $Q_{s}$.  The maximum value of the sum of these probabilities if they result from a single joint distribution is $C_{s}$.  In order to define $p$, suppose Alice and Bob measure their qudits and announce their basis choices, i.e.\ which observable they measured.  In order to guess Bob's result, Alice finds the probability in the set $S_{p}$ corresponding to her measurement result and Bob's basis choice, and guesses that Bob's result is the measurement result for Bob that appears in this probability.  Doing so, she will be correct with probability $p$.  The mutual information between Alice and Bob is found as follows.  Suppose Alice measured $a_{j}$ and Bob measured $b_{k}$.  The mutual information is then given by 
\begin{eqnarray}
I_{ab} & = & \sum_{m,n=0}^{d-1} P(a_{j}=m, b_{k}=n) \nonumber \\
 & & \log_{2} \left[ \frac{P(a_{j}=m, b_{k}=n)}{P(a_{j}=m) P(b_{k}=n)} \right] .
\end{eqnarray}
For the quantum states we are considering, $I_{ab}$ is independent of $j$ and $k$.

Note that the size of the quantum violation of the classical bound decreases with $d$, and the probability of Alice being able to guess the result of Bob's measurement  from her measurement does as well.  However, the mutual information between Alice and Bob increases with $d$.  This is because while the probability of Alice guessing Bob's result is decreasing, the number of alternatives from which she is choosing is increasing.  Thus, Alice and Bob share more information as $d$ increases.

\section*{Acknowledgment}
This research was supported in part by a grant from the Templeton Foundation.

\end{document}